\documentclass[preprint]{vgtc}

\graphicspath{{figures/}{pictures/}{images/}{./}}

\usepackage{times}

\usepackage{tabu}
\usepackage{booktabs}
\usepackage{lipsum}
\usepackage{mwe}

\usepackage{mathptmx}

\usepackage{pgf-umlcd}

\tikzstyle{umlcd style enum}=[rectangle split, rectangle split parts=2,
every text node part/.style={text centered},
draw, minimum height=2em, umlcolor, minimum width=2cm, text width=4cm,
minimum height=1cm, node distance=2cm]
\makeatletter
\newenvironment{enum}[3][]
{
\begin{pgfumlcd@classAndInterfaceCommon}{#1}{#2}{#3}
}
{\node[umlcd style enum, anchor=north] (\pgfumlcd@ClassName) at (\pgfumlcd@ClassPos)
    {$<<$enumeration$>>$ \\ \textbf{\pgfumlcd@ClassName}
\nodepart{second}
\pgfumlcd@ClassAttributes
};
\end{pgfumlcd@classAndInterfaceCommon}
}
\makeatother

\vgtccategory{Research}
\vgtcinsertpkg

\preprinttext{}

\usepackage{float}
\usepackage{listings}

\newfloat{lstfloat}{htbp}{lop}
\floatname{lstfloat}{Listing}
\crefname{lstlisting}{Listing}{Listings}

\usepackage{xspace}
\usepackage{xcolor}
\definecolor{commentgreen}{RGB}{2,112,10}
\definecolor{eminence}{RGB}{108,48,130}
\definecolor{weborange}{RGB}{255,165,0}
\definecolor{frenchplum}{RGB}{129,20,83}
\definecolor{dkgreen}{rgb}{0,0.6,0}
\definecolor{dkblue}{rgb}{0,0,0.6}
\definecolor{gray}{rgb}{0.5,0.5,0.5}
\definecolor{mauve}{rgb}{0.58,0,0.82}
\usepackage{listings}
\newcommand{\code}[1]{\texttt{#1}}

\def\etal{\emph{et~al}.}

\definecolor{ForestGreen}{HTML}{009B55}

\def\rafi{\texttt{RaFI}\xspace}
\def\streami{\texttt{Streami}\xspace}

\title{Streami: An MPI Data-Parallel Library to Compute Field Lines on GPUs}

\author{Stefan Zellmann\thanks{e-mail: zellmann@uni-koeln.de}\\
        \scriptsize University of Cologne
\and Milan Jaros\thanks{e-mail: milan.jaros@vsb.cz}\\
     \scriptsize University of Ostrava
\and Andrea Paris\thanks{e-mail: aparis@nvidia.com}\\
     \scriptsize \centering NVIDIA
\and Ingo Wald\thanks{e-mail: iwald@nvidia.com}\\
     \scriptsize \centering NVIDIA
\and Tatiana von Landesberger\thanks{e-mail: tlandesb@uni-koeln.de}\\
     \scriptsize \centering University of Cologne}

\teaser{
  \centering
  \includegraphics[width=\linewidth]{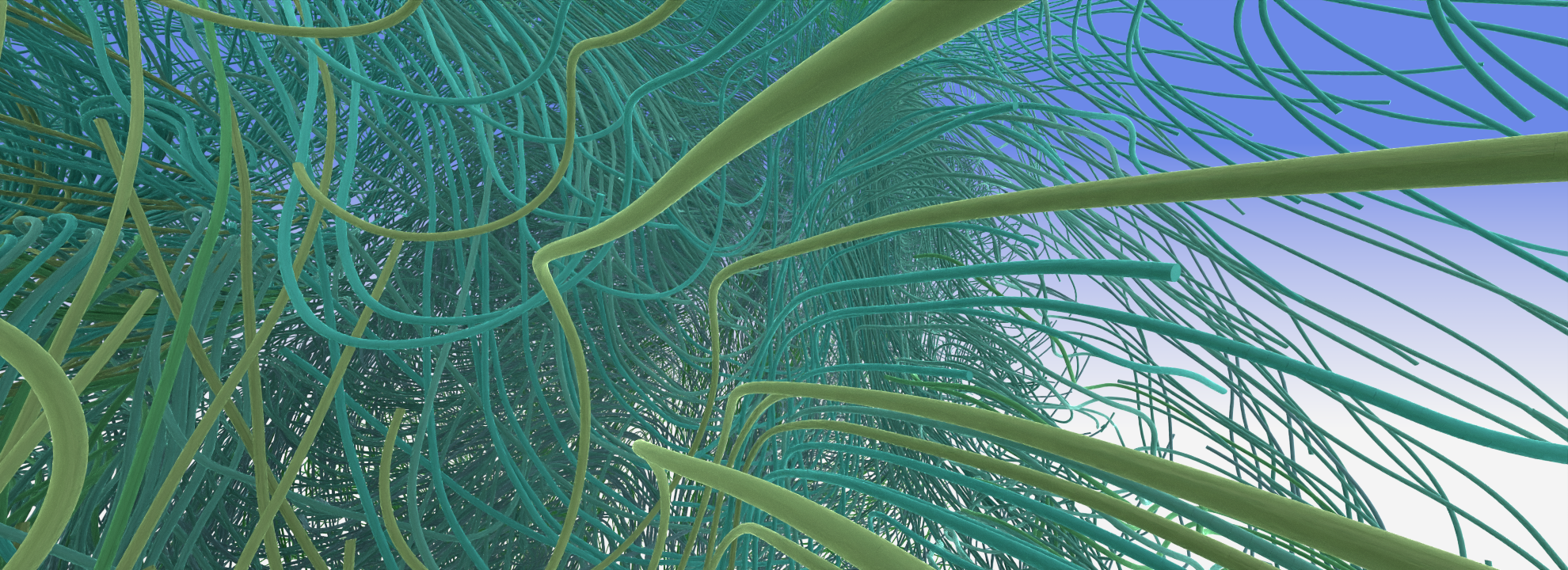}
  \vspace{-1.5em}
  \caption{\label{fig:teaser}
    Magnetohydrodynamics (MHD) compressible turbulence flow visualization with streamlines using \streami.}
}

\abstract{
We present \streami, an extensible GPU-accelerated library for the computation of field lines in fluid flows on high-performance computers.
\streami acts as a thin layer used for both post-hoc or in-situ analysis and can interface with existing MPI applications.
We discuss \streami's application programming interface, key design decisions that led to \streami's high performance and
extensibility, as well as extensions to support different fluid flow field representations.
We also present a sample application for rapid prototyping and interactive seed point placement.
\streami is released under a permissive open-source software license.
}


\nocopyrightspace

\begin{document}

\firstsection{Introduction}

\maketitle

Field lines are important visualization primitives for the analysis of fluid
flows. The different flavors of field lines (e.g., streamlines as shown in
\cref{fig:teaser}, or streaklines used to visualize unsteady flow) are often
used for post-processing computational fluid dynamics (CFD) simulations. Due to
their computational demands, CFD simulations are typically performed on high-performance computing (HPC) systems, which are operated in a distributed memory
fashion and consist of multiple compute nodes, each of which is equipped with
multiple GPUs. While the computations on the individual GPUs are usually
accelerated using GPGPU APIs (e.g., CUDA, HIP, SYCL), communication across compute node boundaries is handled via MPI.
To avoid bottlenecking the application, communication can leverage direct memory access
through MPI extensions (e.g., \emph{CUDA-aware MPI} on NVIDIA systems).

CFD simulations usually run in a data-parallel fashion using domain decomposition, and again, to avoid bottlenecking the application, it is
paramount that post-processing routines such as field line computation
reuse this partitioning to avoid costly data copies. It is usually not feasible
to copy the data to a graphics workstation as a whole to perform further
(post-hoc) analysis. Instead, the visualization is also performed on the HPC
system and, therefore, must also be data-parallel, either by reusing the exact application
context and MPI communicator used by the CFD application (in-situ
visualization), or by transferring the data to a dedicated visualization
cluster (in-transit visualization).

To facilitate these types of visualization modes where the post-processing
predominantly runs on a multi-GPU cluster using CUDA-aware MPI, several
abstractions are required, some of which already have highly performant open source
implementations, and others that are ill-supported by existing tools.
On the in-situ and in-transit side, libraries like
Catalyst2~\cite{catalyst2} or Ascent~\cite{larsen-ascent:2022} implement interfaces and negotiate
between the CFD application and the low-level visualization interface. In the
field of isosurface or volume visualization (targeted at scalar fields), there exist a variety
of applications and software abstractions to implement that low-level
interface without introducing bottlenecks through unnecessary data copies or
context switches. When visualizing large, data-distributed scalar fields, tools
like data-parallel ANARI~\cite{dpanari} or OSPRay~\cite{wald2017ospray} provide
an interface that allows the app to pass the data over in the form that it also
uses internally, and run the visualization algorithm on the given data
distribution, while maintaining high performance through low-level rendering. For
field line computation, such an abstraction does not exist. Instead, these
algorithms are tightly integrated within their respective ecosystem (e.g., VTK~\cite{vtkBook}
or Viskores~\cite{vtkm}).

We present \streami, a simple drop-in library that can efficiently compute
field lines from both steady and unsteady flows given a parallel data
distribution on clusters of multi-GPU systems. \streami sits between the
low-level rendering library and the CFD simulation, typically relying on an in-situ
library to negotiate between the two. \streami can be extended to
accommodate different vector field types using a low-level interface,
alleviating the need for resampling. How exactly the data is distributed
is transparent to the library, provided that the user has access to the spatial
distribution provided by the application. Algorithms like streamline or streakline
computation use the low-level interface and can use functions like particle
tracing kernels or field abstractions as intrinsics, without having to
interface directly with CUDA or MPI. These algorithms present a high-level
interface that is conveniently accessed through a C++ API. We also present
sample applications, including an interactive sample application for seed point
placement based on this API. \streami is open source and published under the
permissive Apache~2.0 software license.

\section{Related Work}
Particle advection is the predominant technique to analyze patterns and
movement in fluids~\cite{gunther-vortex-state-art:2018}. Flow visualization
algorithms based on particle advection perform three main operations: seed
placement, particle updates, and output construction. For field
lines this is dominated by the update step, which is responsible for
particle advection and is itself composed of cell location and interpolation to
evaluate the flow field. We refer to Yenpure
\etal~\cite{yenpure-state-art:2023} for a literature overview on optimization
techniques for these operations.

Because of its high computational cost, the workload of particle advection is
usually distributed across multiple processors. The most common approach is
\emph{parallelization over data} (POD)~\cite{wang-parallelize-over-data:2025},
often using simple a block partition as data distribution. Hybrid approaches
were also proposed that distribute both the seed points (parallelization over
seeds), and the data on demand~\cite{pugmire:2009}. POD is known to be prone to load
imbalances~\cite{peterka-loadbalance,load-balance-streamlines}, which can be mitigated if the user
is in control of the data partitioning. Other factors that affect
performance, such as the ping-pong effect described by Wang
\etal~\cite{wang-parallelize-over-data:2025} are also related to the data
partition and potential halo regions. Since these challenges are so specific
to the data and its topology, a library such as ours should leave these tasks of
how the data is partitioned, how neighbor queries are performed to locate partitions
on other ranks, or how connectivity and overlap or halo regions across ranks are set up,
to the user rather than imposing its own data partitioning scheme.

Another customization point to the library is field topology and cell
location. Simple vector fields consist of voxel grids, where cell
location is a simple array access. Fluid simulations often involve more
complex mesh geometry~\cite{tri-quad-streamlines} or hierarchical grid data
structures~\cite{zellmann-cise:2022} though, so that tree traversal or similar
operations are used for cell location. Tools like ParaView~\cite{paraview} or
VisIt~\cite{visit} are restricted to a number of different field types,
and to visualize custom flow fields, resampling or remapping is necessary,
often resulting in inferior quality or excessive memory use.
Performance-portable frameworks for field line computation have been presented
for many-core architectures~\cite{pugmire-vtkm-advection:2018}, but a high
performance library for multi-node/multi-GPU systems using MPI, as contributed by
\streami, is currently missing from the state of the art.

\section{Software Library Design}
\streami is composed of a low- and a high-level API. The low-level API is
implemented in CUDA/C++, while the high-level API is written in object-oriented
C++. The low-level interface provides the building blocks for seed point
placement, particle advection, and output assembly; the high-level interface
combines those building blocks to form flow visualization algorithms for field
line computation. For optimal performance, the low-level interface can be
extended with new flow field types via CUDA/C++ templates. We describe the APIs
in detail in the remainder of this section.

\subsection{Low-Level Interface}
  \tikzset{
  every picture/.append style={
    transform shape,
    scale=0.7
  }
}
\begin{figure}[tb]
  \centering
\begin{tikzpicture}
      \begin{class}[text width=6cm]{VecField}{0,0}
        \attribute{worldBounds : AABB}
        \operation{sample(float3 position, float3 \&retval) : bool}
        \operation{destinationID(float3 position) : int}
      \end{class}
      \begin{class}[text width=4cm]{MacroCell}{6.5,0}
        \attribute{ID : int}
        \attribute{bounds : AABB}
        \attribute{domain : AABB}
      \end{class}
      \begin{class}[text width=4cm]{RankInfo}{0,-3}
        \attribute{mpiRank : int}
        \attribute{mpiCommSize : int}
      \end{class}
      \composition{VecField}{mc}{1}{MacroCell}
      \composition{VecField}{ri}{1}{RankInfo}
\end{tikzpicture}
\caption{\label{fig:device-classes}
Class diagram for the distributed vector field abstraction on one GPU. Using
the compile-time polymorphic function \code{VecField::destinationID()}, CUDA
kernels can determine the MPI rank to which particles exiting this rank's domain must be transferred.
}
\vspace{-1em}
\end{figure}
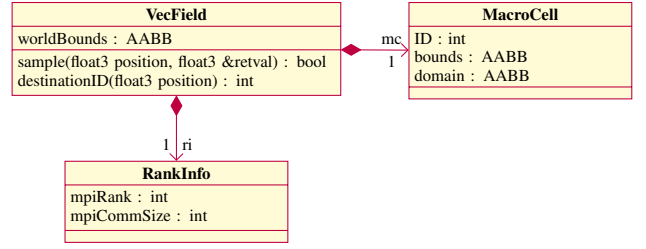
The main abstraction of \streami's low-level interface is the \emph{vector
field}. The vector field consists of a host and device class enabling the implementation of basic control flow. While the host class is primarily responsible for
data handling, the
device class (\cref{fig:device-classes}) implements the main utilities that are
used by the particle tracing algorithms in CUDA. The device class can be
extended with user-defined vector field types at compile time using template
instantiation mechanisms. We resort to compile time-polymorphic execution as
these operations will be performed in inner loops on the GPU and are critical
for performance. An instance of that class is present on each GPU used by the
MPI app.

\subsubsection{Fluid Field Sampling and Particle Forwarding}
The core functions provided by the low-level interface that must
be implemented for every vector field type are
\code{VecField::sample()} and \code{VecField::destinationID()}. The function
\code{sample()} reconstructs the data at a given position in space. It returns
the value of the vector field as an output parameter and \code{true} as function
return value upon success, or \code{false} as the return value if the field is
undefined at that position---or a position was sampled for which the data is
present on another GPU. The function \code{destinationID()} takes a 3D position
in space and computes the MPI rank of the GPU holding the data for that
position. This operation is tied to the field type implementation and can be
redefined by flow field extensions. The ID is used to determine the rank the
particle is transferred to next. \streami assumes that there is one MPI
process per GPU.

\begin{figure}[t]
\centering
  \includegraphics[width=0.49\columnwidth]{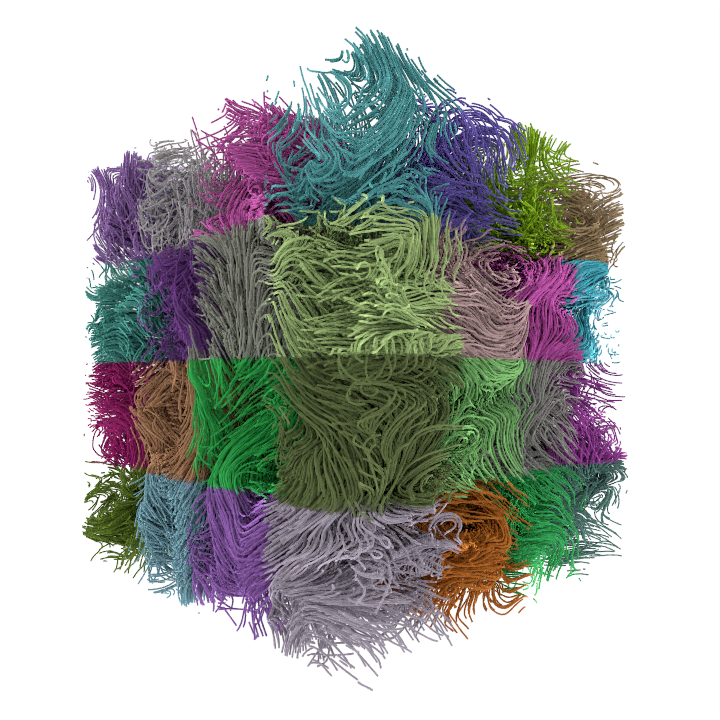}
  \includegraphics[width=0.49\columnwidth]{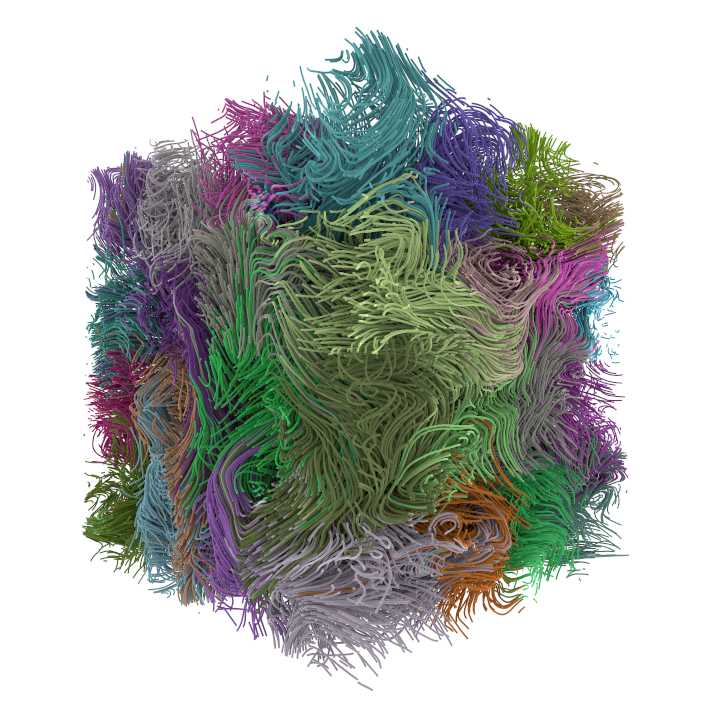}
\vspace{-1.5em}
\caption{\label{fig:rank}
Turbulent flow structured field advection, data distributed to 64 MPI ranks.
Left: color by rank assignment during advection. Right: color by birth rank,
showing the particle movements across MPI processes.
}

\vspace{-1.5em}
\end{figure}

The abstraction assumes that each partition of the data resident on a GPU is
represented by a \emph{macrocell}. The combination of vector field \emph{and}
macrocell that is referenced by the field instance uniquely defines the part of
the data present on one GPU. The macrocell can also be uniquely identified by
its MPI rank, i.e., \code{destinationID()} must provide a unique mapping from
positions in space to flat integer IDs. A possible representation is that the
field organizes the data in a uniform rectangular grid, each GPU is assigned
one cell of that grid, and the CUDA device function \code{destinationID()}
performs a simple offset computation given the projected cell position and
grid dimensions. Other representations are easy to support though; it would, e.g., also
be perfectly valid that the implementation for this customization point
performed some kind of (on-device) tree traversal to locate the macrocell.
The only requirement is that the underlying
spatial partition can identify axis-aligned bounding boxes (AABB) associated
with the GPU's portion of the data. Macrocells also store the \emph{domain}
bounds which include a user-defined halo region for processing at the node
boundaries. Example rank assignments and particle movement based on birth rank
are shown in \cref{fig:rank}, for uniform grid macrocells, using the
magnetohydrodynamics compressible turbulence data
set~\cite{burkhart2020catalogue,ohana2024well}.

\subsubsection{Particle Advection and GPU-to-GPU Communication}
\begin{lstfloat}
\begin{lstlisting}[%style=cppstyle,
    caption={GPU particle advection kernel in CUDA-inspired pseudocode. The
    kernel uses the device vector field abstraction to sample the flow
    field, computes new positions using Runge-Kutta, and transfers
    particles between GPUs based on those position.
},
label={lst:update-kernel}]
__global__ void update(VecField field) {
  int particleID = cudaThreadIndex();
  Particle p = getIncoming(particleID);
  float3 P0 = p.position;

  // compute new position using Runge-Kutta
  float3 k1, k2, k3, k4;
  if (!field.sample(P0,k1)) return;
  float3 P1 = P0+k1*0.5f;
  if (!field.sample(P1,k2)) return;
  float3 P2 = P0+k2*0.5f;
  if (!field.sample(P2,k3)) return;
  float3 P3 = P0+k3;
  if (!field.sample(P3,k4)) return;
  p.position = P0+1/6*(k1+2*k2+2*k3+k4);

  // compute and emit to new rank
  int dest = field.destinationID(P);
  emitOutgoing(p,dest);
}
\end{lstlisting}
\end{lstfloat}
Given the two intrinsic functions for sampling and destination rank computation,
particle advection can be implemented using a CUDA GPU kernel
(\cref{lst:update-kernel}). The kernel takes the field as input, and is invoked
for all the particles the GPU is responsible for. These invocations can run in
parallel on each GPU. The execution model assumes that each GPU knows all the
particles that are advected, but that is a detail only the high-level interface
described below is aware of. To the MPI process and local GPU context, all that
matters is that there is an array with a number of particles given by
their position, and a unique ID so we can track them across devices.
A vector field must be given that can be used to advect those particles and
transfer them to their new destination device. The kernel hence executes one
CUDA thread per particle and advects it through space using the Runge-Kutta
algorithm by sampling offset vectors from the field. Given the new particle
position, the kernel computes the device ID and enqueues it so the host can
forward it to its new destination. The destination can be the same GPU
from which the particle originated.

Particle exchange is implemented using the optimized CUDA library
\rafi~\cite{rafi}, whose original task was to forward rays across ranks in a
data-parallel MPI ray tracer. We use this library as a black-box treating
particles as rays. \rafi internally maintains our particle buffers, including
the incoming particle array, and provides CUDA device functions to read from
(\code{getIncoming(particleID)}) and write to
(\code{emitOutgoing(particle,destID)}) those buffers. \code{emitOutgoing()} does
not directly initiate communication, but buffers the particles. The entire
buffer is transferred using a collective all-to-all call provided by \rafi that
internally exchanges the local buffers using CUDA-aware MPI.

These abstractions consisting of kernels, MPI particle forwarding, and
vector/fluid field implementations form the low-level interface of \streami. On
this level the abstractions use CUDA/C++ templates to achieve speed-of-light
performance. For practical reasons the low-level interface also implements a
number of convenience functions, most of which manifest as CUDA kernels. Those
include, e.g., generating uniformly distributed seed points given simple
control shapes such as planes or lines, etc.

\subsection{High-Level Library Interface}
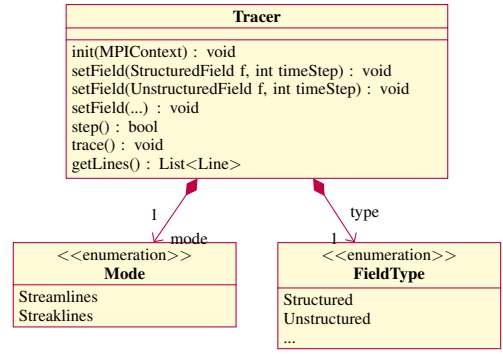
\begin{figure}[tb]
  \centering
\begin{tikzpicture}
    \begin{enum}[text width=3cm]{Mode}{0,-4.5}
       \attribute{Streamlines}
       \attribute{Streaklines}
     \end{enum}
    \begin{enum}[text width=3cm]{FieldType}{5,-4.5}
       \attribute{Structured}
       \attribute{Unstructured}
       \attribute{...}
     \end{enum}
      \begin{class}[text width=7cm]{Tracer}{2.5,0}
        \operation{init(MPIContext) : void}
        \operation{setField(StructuredField f, int timeStep) : void}
        \operation{setField(UnstructuredField f, int timeStep) : void}
        \operation{setField(...) : void}
        \operation{step() : bool}
        \operation{trace() : void}
        \operation{getLines() : List$<$Line$>$}
      \end{class}
      \composition{Tracer}{mode}{1}{Mode}
      \composition{Tracer}{type}{1}{FieldType}
\end{tikzpicture}
\caption{\label{fig:tracer}
Tracer class provided by the high-level C++ interface.
}
\vspace{-1em}
\end{figure}
With the low-level abstraction it is easy to implement different field line
computation algorithms. For steady-state simulations, streamlines are effective
as the vector fields that do not change over time. This algorithm
consecutively calls the low-level \code{update()} kernel,
interleaved with particle forwarding using \rafi. Unsteady flow with
streaklines is a simple extension that, on an abstract level, periodically adds
seed points and connects them in a different order than it does with
stremalines. We provide a high-level \code{Tracer} class
that implements these algorithms, shown in \cref{fig:tracer}.

The \code{Tracer} class is initialized with an app-provided MPI context so that
each process has its own instance of the tracer class. All interprocess
communication is immediately offloaded to \rafi. Fields are assembled by the
app by using their compile-time polymorphic types. This allows the app to
provide the data in whatever is the most efficient way, with the only condition
being that the data is compatible with one of the given field types. The
high-level API is designed to maintain one field per time-step, which is relevant
for unsteady-state simulations. For simple field types this is a reasonable
choice regarding performance, memory consumption, and simplicity in
implementation. For fields that consist of polyhedral or other
unstructured element types, this design choice can however lead to increased
memory consumption if only the data, but not the topology changes
over time. We for now prefer the simpler abstraction but may change the design
in the future if the need arises. Once the tracer is initialized properly, the
app can call \code{step()} to perform a particle advection step, or \code{trace()}
to step until a maximum number of steps are reached or no particles are left to
advect.

\begin{lstfloat}
\begin{lstlisting}[%style=cppstyle,
    caption={C++ pseudocode for the \code{step} function to advect particles.
},
label={lst:step}]
bool Tracer::step() {
  if (mode == Streaklines) {
    simTime += dt; // advance simulation time
    simStep = (int)simTime;
  }
  auto field = fields[simStep];

  // call low-level update function
  if (type == Structured)
    call_update_Structured(field.asStructured());
  else if (type == Unstructured)
    call_update_Unstructured(field.asUnstructured());
  else if (...) ... // other field type extensions

  // unsteady flow: enter more particles
  if (mode == Streaklines) call_generateNewParticles();

  assembleOutput(); // make lines/curves

  // use RaFI to exchange particles between MPI processes
  auto fwdResult = rafi_forwardParticles();
  return fwdResult.numActiveAcrossAllRanks != 0;
}
\end{lstlisting}
\end{lstfloat}
\code{step()} and \code{trace()} are also responsible for advancing the
simulation time step for unsteady flow. In the case of streamlines and steady
flow, particles are generated only once, on advection step $0$, while for
streaklines a new batch of particles is generated per advection and simulation
time step. This update logic and particle buffer management is hidden inside
the tracer class, which simply calls the respective low-level functions for
particle generation and updates. Pseudocode for the step function is shown in
\cref{lst:step}.

Finally, the field lines need to be assembled so the user can viualize them.
This is accomplished by calling \code{Tracer::getLines()} which implements this
logic based on the field line type chosen. We currently focus on streamlines
and streaklines. Other representations (e.g., path lines) would be easy to
implement within the framework.

\subsection{Fluid Field Extensions}
The field abstraction allows us to extend the library with different field types.
On the host side this requires implementing an abstract base class, while on the
device we resort to compile time polymorphism for performance reason; i.e.,
different fields and acceleration structures must be added to the library
directly and not through a plug-in mechanism. Semantically, fields are responsible
for two operations: for sampling the fluid flow at a position in 3D space, and to
find the spatial partition and GPU/MPI rank that belongs to
that position in space; i.e., the topology of the spatial
subdivision is also hardcoded into the vector field type. A field using voxels to
store the data and a uniform grid of macrocells for the data distribution would
have a different implementation than a voxel field using a kd-tree for space
partitioning. This is aided by C++ template mechanisms though, so that
implementations can share code, e.g., for sampling the field while using
different spatial data structures.

The combinations we currently provide are a uniform grid storing voxels of type
\code{float3}, and an unstructured grid type supporting tetrahedra, pyramids,
prims, and hexahedra. The structured field uses a CUDA global memory array for the
voxels. Sampling is done by computing offsets into that array. The unstructured
field uses a GPU-accelerated bounding volume hierarchy, realized with
cuBQL~\cite{cubql}, and point containment tests for sample location. Both field
types use a uniform grid of macrocells as distributed spatial subdivision whose
dimensions can be configured at runtime to match them to those used by the app.
Further extensions remain future work but are easy to add using our framework.

\section{Sample Apps}
\begin{figure}[t]
\centering
  \includegraphics[width=0.49\columnwidth]{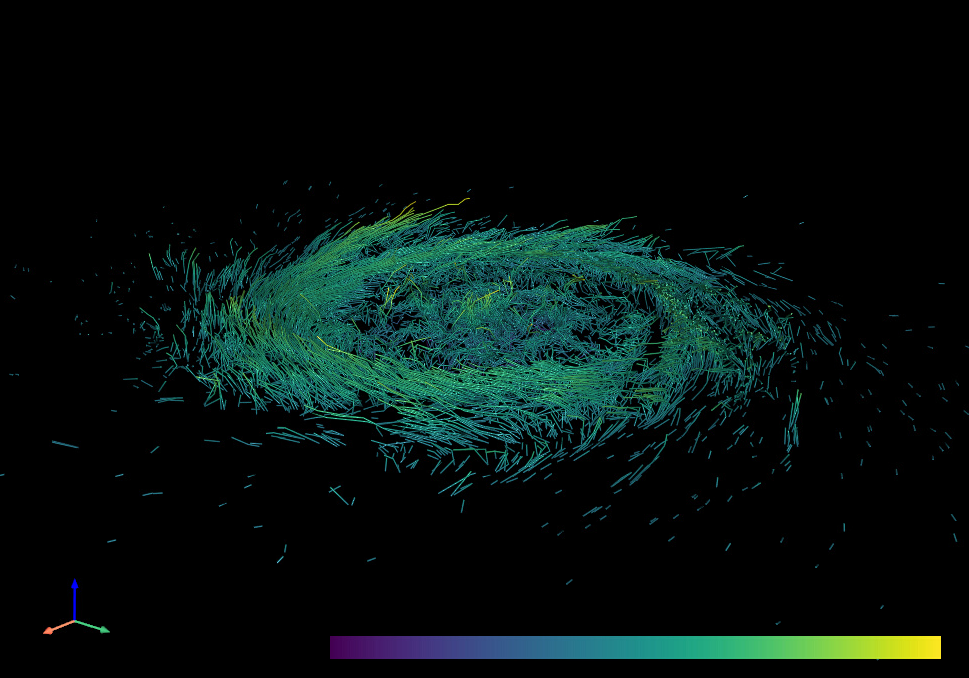}
  \includegraphics[width=0.49\columnwidth]{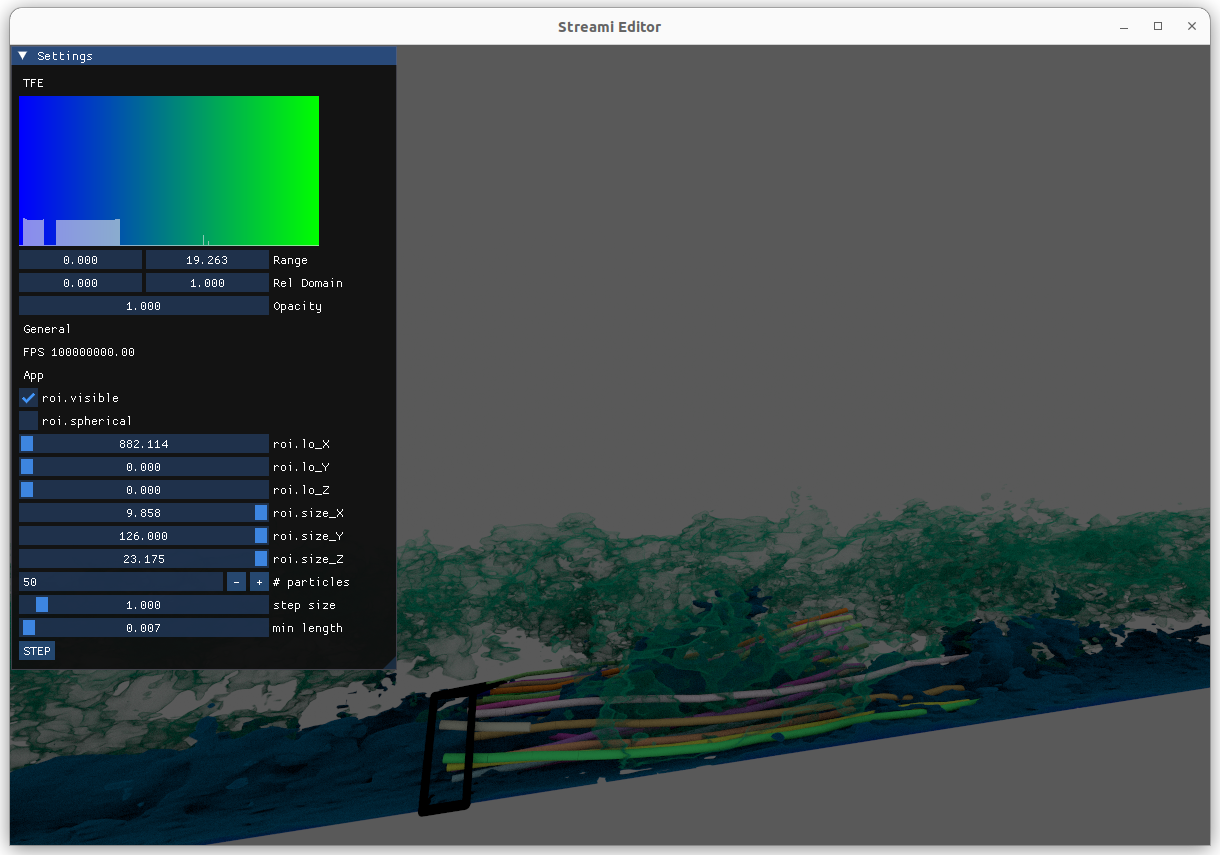}
\vspace{-1em}
\caption{\label{fig:apps}
Sample apps using \streami. Left: streamlines generated with the command line
app, for the astrophysics data set from~\cite{Wissing2023}, and visualized with
PyVista. Right: Interactive sample app using \streami with interactive seed
point placement and volume rendering overlay of the turbulence field, using the
wind farm data set from~\cite{windfarm}.
}
\vspace{-1em}
\end{figure}
We implemented two sample apps that use \streami: a command line app that
writes its output to files, and an interactive app for
exploration (see \cref{fig:apps}). The apps are responsible for creating an MPI
context, a \code{Tracer} class instance using the C++ interface, and to call
\code{Tracer::step()} or \code{Tracer::trace()} for an appropriate number of
times. The command line app produces wavefront obj files containing line
segments as output. Before creating those files, the main rank gathers all
advected particles across all time steps from all MPI ranks and writes them to
a joint obj file. The files can then be opened with an external 3D application.
The interactive app provides a user interface to set up and interactively
adjust seed point regions to rapidly explore the data. It uses ANARI for 3D
rendering and also includes a volume rendering of the input field converted to
scalars given by the vector norms to improve explorability.

\section{Results and Discussion}
\begin{figure}[t]
\centering
  \includegraphics[width=0.49\columnwidth]{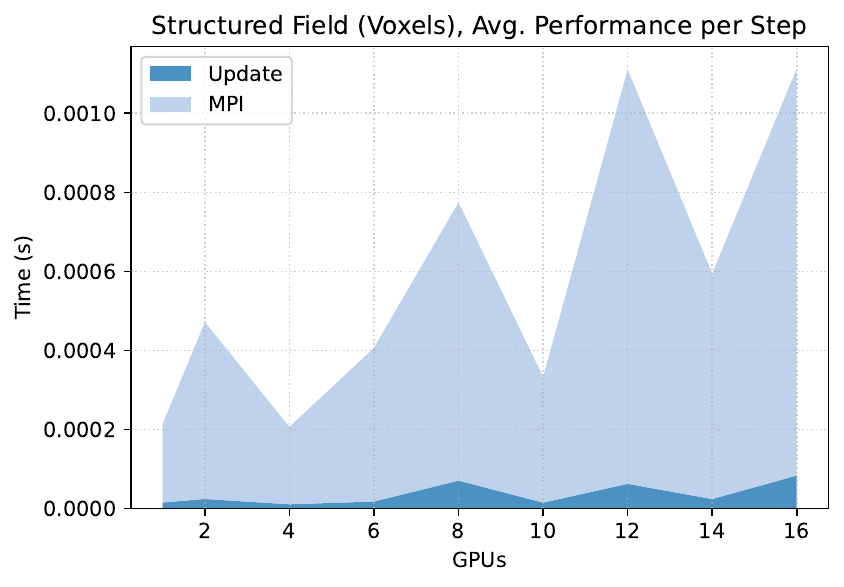}
  \includegraphics[width=0.49\columnwidth]{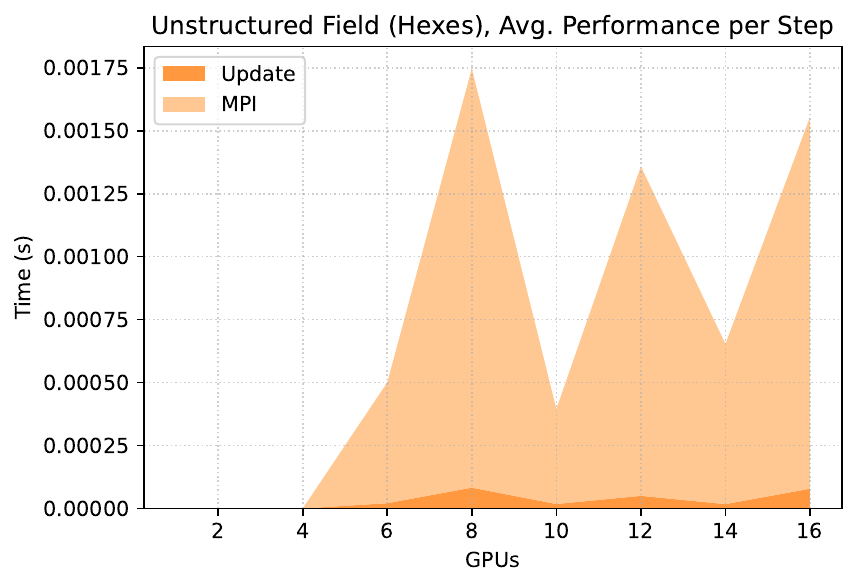}
\vspace{-1em}
\caption{\label{fig:results}
Performance on up to 16$\times$ A100 GPUs (two nodes with eight GPUs), for
the galaxy data set~\cite{Wissing2023} ($1K^3$ voxels). We report timings for
tracing 100K particles, averaged over 10K advection steps. (Unstructured:
out-of-memory for 1, 2, and 4 GPUs.)
}
\vspace{-1em}
\end{figure}
The main contribution of \streami is its API that is designed for simple
integration into existing HPC post-processing pipelines. As a qualitative
result, this is hard to evaluate. A core property of \streami is its high
performance on GPU clusters, but evaluating that is also anecdotal since a typical
workload will vary in data size, the number of particles advected, the shape and
locality of seed point regions, and many other factors. As proof that \streami
operates within reasonable parameters, we include a benchmark showing
performance numbers for the local advection phase, and for the parallel
particle forwarding phase using CUDA-aware MPI, in \cref{fig:results}. For
those, we use the astrophysics data set by Wissing and Shen~\cite{Wissing2023}
that is given as a structured volume. To obtain results for unstructured data
we also converted the data set to an unstructured flow field so that all voxels
become hexahedra. The results show that \streami is mainly limited by MPI
communication; single advection steps for this workload take on the order of
1-2~ms. Finally, we also released \streami on GitHub (\url{https://github.com/szellmann/streami}).

\section{Conclusion}
We presented \streami, a high performance, data parallel library for field line
computation using CUDA and MPI. We discussed \streami's low- and
high-level APIs as well as the design decisions that led to them. \streami is
designed to interoperate with HPC simulation apps to provide efficient field
line computation in in-situ and in-transit scenarios. We achieve this using an
extensible API that allows for efficient reuse of the domain decomposition
between \streami and the app, by that adding minimal overhead to the
interactive post or in-situ processing phase of the simulation.

\acknowledgments{
Funded by the Deutsche Forschungsgemeinschaft (DFG, German Research Foundation) under Germany’s Excellence Strategy EXC 3037 – 533607693– Unser dynamisches Universum.
Additionally, this work was funded by the German Federal Ministry of Research, Technology and Space under the funding code~01LK2204A. The responsibility for the content of this publication lies with the authors.
This work was also supported by the Ministry of Education, Youth and Sports of the Czech Republic through the e-INFRA CZ (ID:90254).}

\bibliographystyle{abbrv-doi}

\bibliography{bibliography}
\end{document}